\documentclass[fleqn,usenatbib]{mnras}

\usepackage[T1]{fontenc}
\usepackage{ae,aecompl}

\usepackage{graphicx}	
\usepackage{amsmath}	
\usepackage{amssymb}	

\newcommand{\delr}{\frac{\partial}{\partial r}}

\newcommand{\delf}[2]{\ensuremath{\frac{\partial #1}{\partial #2}}}
\newcommand{\df}[2]{\ensuremath{\frac{d #1}{d #2}}}

\title[From Birth to Death of Protoplanetary Disks]{From Birth to Death of Protoplanetary Disks: Modeling Their Formation, Evolution, and Dispersal}

\author[S.S. Kimura, M. Kunitomo, S.Z. Takahashi]{
Shigeo S. Kimura$^{1,2}$\thanks{E-mail: shigeo@astr.tohoku.ac.jp}, 
Masanobu Kunitomo$^{3}$, 
and Sanemichi Z. Takahashi$^{2}$
\\
$^1$Frontier Research Institute for Interdisciplinary Sciences, Tohoku University, Sendai 980-8578, Japan\\
$^2$Astronomical Institute, Tohoku University, Sendai 980-8578, Japan\\
$^3$Department of Physics, Nagoya University, Nagoya, Aichi 464-8602, Japan
}

\date{Accepted XXX. Received YYY; in original form ZZZ}

\pubyear{2016}

\begin{document}
\label{firstpage}
\pagerange{\pageref{firstpage}--\pageref{lastpage}}
\maketitle

\begin{abstract}
Formation, evolution, and dispersal processes of protoplanetary disks are investigated and the disk lifetime is estimated. Gravitational collapse of a pre-stellar core forms both a central star and a protoplanetary disk. The central star grows by accretion from the disk, and irradiation by the central star heats up the disk and generates thermal wind, which results in the disk dispersal. Using the one-dimensional diffusion equation, we numerically calculate the evolution of protoplanetary disks. To calculate the disk evolution from formation to dispersal, we add source and sink terms that represent gas accretion from pre-stellar cores and photoevaporation, respectively. We find that the disk lifetimes of typical pre-stellar cores are around 2--4 million years (Myr). A pre-stellar core with high angular momentum forms a larger disk whose lifetime is long, while a disk around a X-ray luminous star has a short lifetime. Integrating the disk lifetimes under various mass and angular velocity of pre-stellar cores and X-ray luminosities of young stellar objects, we obtain disk fraction at a given stellar age and mean lifetime of the disks. Our model indicates that the mean lifetime of protoplanetary disks is 3.7 Myr, which is consistent with the observational estimate from young stellar clusters. We also find that the dispersion of X-ray luminosity is needed to reproduce the observed disk fraction.
\end{abstract}




\section{Introduction}

Protoplanetary disks are the birth place of planets. Revealing their formation, evolution, and dispersal processes is essential to understand planet formation. 

The system of a protostar and a protoplanetary disk is formed by gravitational collapse of a rotating pre-stellar core in a molecular cloud. Observations provide us the statistical information of mass  \citep[e.g.][]{ADW14a} and velocity gradient \citep[e.g.][]{CBM02a} of pre-stellar cores. Using these information as an initial condition, three-dimensional numerical simulations of collapsing pre-stellar cores have unveiled the formation process of protostars and protoplanetary disks \cite[e.g.][]{MIM10a,Inu12a,TTM15a}. 
However, such three-dimensional simulations cannot follow the long-term evolution of the system of a protostar and a disk because of their expensive computational cost. 

Most of gas in a pre-stellar core accretes onto a disk because of the centrifugal force. For the early evolutionary phase of this system, the disk is gravitationally unstable because the disk is more massive than the protostar \citep[e.g.][]{VB06a,MIM10a,IMM10a}. During this phase, gravitational torque by spiral arms formed in the disk efficiently transports the angular momentum, and the disk supplies the mass and entropy to the protostar through accretion, which affects evolutionary tracks of protostars \citep{BCG09a,HOK11a,BVC12a}. After most of the gas accretes onto the disk, the protostar mass continues to increase whereas the disk mass decreases because of mass accretion from the disk to the protostar, so that the disk becomes gravitationally stable. Then, the turbulent stress caused by magneto-rotational instability (MRI) plays an important role for disk evolution \citep{bh91}. 
After the accretion timescale of the protostar becomes shorter than its thermal timescale, it starts the Kelvin-Helmholtz contraction \citep[pre-main sequence,][]{Hay61a}. 

The protoplanetary disks are observed as infrared excesses in the spectroscopic observation of young stellar objects (YSOs). Infrared observations of young star clusters give the relation between the cluster age and disk fraction. These data suggest that protoplanetary disks disappear in several million years (Myr) after the formation of pre-main sequence stars \citep[e.g.][]{YKT10a,Mama11a}. 
The widely accepted picture of disk dispersal is the combination of viscous accretion onto the star with the photoevaporation that is the thermally-driven wind induced by the irradiation of ultraviolet (UV) and X-rays \citep[e.g.][]{CGS01a}.
Previous theoretical studies claim that the combination of photoevaporation and viscous evolution can explain the observed disk lifetime \citep[e.g.][]{AA09a,GDH09a,OEC10a,BHZ13a}. However, the formation process of disk has been ignored in these studies except for \cite{BHZ13a}.

The formation and dispersal processes of protoplanetary disks have been independently investigated so far. 
However, the disk evolution depends on the initial surface density profile, which is determined by the property of the pre-stellar core. Therefore, to obtain the reliable disk lifetime, the initial disk profile should be given by the model that takes account of collapse of a pre-stellar core. Constructing such a theoretical model enables us to bridge the observations of pre-stellar cores and disk lifetime.
\citet{BHZ13a} construct a model including the formation, evolution, and dispersal of the disk consistently. They calculate the evolution of the disk formed from pre-stellar cores of different angular velocities, and their model seems to be consistent with the disk lifetime estimated by observations. However, they do not consider the mass variation of pre-stellar cores, which is likely to affect the disk fractions estimated from observations of young star clusters. 

In this paper, we calculate the formation, evolution, and dispersal processes of the protoplanetary disks. We start calculations from pre-stellar cores of various mass, and calculate the disk evolution with photoevaporation by X-rays until the disk disperses. We also estimate the disk fraction at a given stellar age and the mean lifetime of disks and compare it to the observations, which is the first attempt that bridges the observations of pre-stellar cores and disk lifetime. This paper is organized as follows. We describe a model of disk formation, evolution of a star and a disk, and photoevaporation in Section \ref{sec:model}. The results of our calculations are represented in Section \ref{sec:results}, and Section \ref{sec:summary} is devoted to summary.

 \section{Model} \label{sec:model}
 
We construct the model including disk formation from a pre-stellar core, co-evolution of a star-disk system, and photoevaporation by X-rays from the central star. To see a wide range of parameter space based on observations of pre-stellar cores, we use some simplifications compared to \citet{BHZ13a}. 

\subsection{Disk evolution}

We solve the one-dimensional diffusion equation to calculate the evolution of protoplanetary disks. We use the cylindrical coordinate ($r,~\phi,~z$), assume the axisymmetric flow, and integrate the fluid equations in vertical direction. Then, using the equations of mass and angular momentum conservation, we obtain the diffusion equation \citep[e.g.][]{pri81},
 \begin{equation}
  \delf{\Sigma}{t} = - \frac 1 r \delr 
\left[{1 \over (\partial j/\partial r) }
 \delr \left( r^3 \nu \Sigma \df{\Omega}{r}\right)\right]
 + \dot \Sigma_{\rm inf} - \dot \Sigma_{\rm PE}, \label{eq:diffusion}
 \end{equation}
 where $\Sigma$ is the surface density, $j$ is the specific angular momentum, $\Omega$ is the angular velocity, and $\nu$ is the kinetic viscosity.  In addition to the viscous evolution, we consider the mass loss term by photoevaporation wind, $\dot \Sigma_{\rm PE}$, and the mass source term by infall from a pre-stellar core, $\dot \Sigma_{\rm inf}$, assuming that the specific angular momenta of the infall and wind are the same as those of the disk. We assume the hydrostatic equilibrium in the vertical direction, and then, the scale height is given as $H=c_{\rm s}/\Omega$, where $c_{\rm s}$ is the sound speed. 
We write $c_{\rm s}^2= k_{\rm B}T_{\rm c}/(\mu m_{\rm H})$, where $k_{\rm B}$ is the Boltzmann constant, $T_{\rm c}$ is the temperature of disk mid-plane, $\mu=2.35$ is the mean molecular weight, and $m_{\rm H}$ is the mass of hydrogen atom. 

In Equation (\ref{eq:diffusion}), we assume the centrifugal balance, $j^3/r^2 = \partial \Phi/\partial r$ ($\Phi$ is the gravitational potential). 
The self-gravity of a disk should be included, since it is comparable to the stellar gravity in our calculations. 
We approximately write the gravitational potential as the spherically symmetric formula, 
 $\Phi=GM(r)/r$, where $G$ is the gravitational constant and $M(r)=M_*+\int_0^r 2\pi r' \Sigma(r') dr'$ is the enclosed mass of the star and the disk inside the radius $r$ ($M_*$ is the mass of the central star). Then, the specific angular momentum and the angular velocity are written as $j=\sqrt{G M(r) r}$ and $\Omega=j/r^2=\sqrt{GM(r)/r^3}$, respectively. 

We use the alpha prescription for viscosity $\nu=\alpha c_{\rm s}^2/\Omega$ \citep{ss73}. In protoplanetary disks, the gravitational torque and/or the torque due to magneto-rotational instability (MRI) are responsible for the angular momentum transport, which gives $\alpha=\max\left(\alpha_{\rm g},~\alpha_{\rm MRI}\right)$, where $\alpha_{\rm g}$ and $\alpha_{\rm MRI}$ represent the torque due to self-gravity and MRI turbulence, respectively. Since the gravitational torque depends on how the disk is gravitationally unstable, $\alpha_{\rm g}$ is often given as a function of Toomre's $Q$ parameter $Q=c_{\rm s}\kappa_{\rm epi}/(\pi G \Sigma) $, where $\kappa_{\rm epi}=\sqrt{(dj^2/dr)/r^3}$ is epicyclic frequency \citep{Too64a}. We use $\alpha_{\rm g}=\exp(-Q^4)$ according to \citet{ZHG10a} \citep[see also][]{TIM13a}. For the turbulent viscosity by MRI, we use a constant value of alpha, $\alpha_{\rm MRI}=0.01$. In reality, the strength of MRI turbulence depends on the resistivity, so that $\alpha_{\rm MRI}$ is expressed as a function of ionization degree \citep{Gam96a,OH11a}. Since this makes disk evolution complicated, we ignore the effect of resistivity as a first step study.


Since the thermal timescale is much shorter than the viscous timescale in the usual protoplanetary disks, we simply determine the temperature from the balance between three heating processes and a radiative cooling. As the heating processes, we consider viscous heating, accretion from the pre-stellar core, and irradiation from the central star and ambient molecular cloud. The viscous heating rate is 
\begin{equation}
 \Gamma_{\rm vis}= \frac 9 4 \nu \Sigma \left(r \delf{\Omega}{r}\right)^2.
\end{equation}
We assume that the specific internal energy brought by infalling material has the same value as the specific kinetic energy of the local Keplerian motion, in which the heating rate by infalling material from the pre-stellar core is written as 
\begin{equation}
 \Gamma_{\rm inf}= \frac 1 2 \dot \Sigma_{\rm inf} v_{\rm K}^2,
\end{equation}
where we consider the rest of energy brought from the pre-stellar core is radiated away. 
The irradiation flux, $\sigma_{\rm SB} T_{\rm irr,*}^4$, due to the central star of luminosity $L_*$ is estimated to be \citep[e.g.][]{MG04a}
\begin{equation}
 \sigma_{\rm SB} T_{\rm irr,*}^4= \frac {L_*} {4\pi r^2} \frac H r \left(\delf{\ln H}{\ln r}-1\right),
\end{equation}
where $\sigma_{\rm SB}$ is the Stephan-Boltzmann constant, $T_{\rm irr,*}$ is the temperature determined only by the irradiation from the central star, and $d\ln H/d \ln r - 1$ is the shielding factor. We fix $d\ln H/d\ln r = 9/7$ \citep{HG05a} to avoid the numerical oscillation, such as thermal instability (see below). 
The irradiation heating rate is represented as \citep[e.g.,][]{MG04a},
\begin{equation}
 \Gamma_{\rm irr}= {8\sigma_{\rm SB} (T_{\rm irr,*}^4 + T_{\rm amb}^4)\over \left(\tau/2+1/\sqrt 3 + 1/(3\tau)\right)},
\end{equation}
where $T_{\rm amb}$ is the temperature of ambient gas in molecular cloud and $\tau=\kappa \Sigma/2$ is the optical depth ($\kappa$ is the opacity). 

The radiation cooling rate is represented as \citep{Hub90a,MG04a}
\begin{equation}
 \Lambda_{\rm rad}=  {8\sigma_{\rm SB}  T_{\rm c}^4\over3\left(\tau/2+1/\sqrt 3 + 1/(3\tau)\right)},
\end{equation}
where $T_{\rm c}$ is the temperature at disk midplane. We calculate the equilibrium temperature of each process, $T_{\rm vis},~T_{\rm inf},$ and $T_{\rm irr}$ from $\Gamma_{\rm vis}=\Lambda_{\rm rad}$, $\Gamma_{\rm inf}=\Lambda_{\rm rad}$, and $\Gamma_{\rm irr}=\Lambda_{\rm rad}$, respectively, and estimate the disk temperature to be
\begin{equation}
 T_{\rm c}^4=T_{\rm vis}^4 + T_{\rm inf}^4 + T_{\rm irr}^4.
\end{equation}
We note that in our model, the disk temperature is usually determined by one dominant process of the three heating processes. In this situation, error is small for the case with the sum of the three temperatures, rather than the average. In reality, $\tau$ depends on the temperature, but for simplicity, we use the optical depth $\tau$ calculated in the previous time step when calculating $T_{\rm vis}$ and $T_{\rm inf}$ ($T_{\rm irr}$ is independent of $\tau$). Opacity in protoplanetary disk is given in \citet{ZHG09a} as a useful fitting formula. This opacity induces thermal instability for optically thick disks for $T_{\rm c}\gtrsim 1500$ K \citep{KT12a}, which leads to the variable mass accretion onto protostar \citep{BL94a,ZHG10a,OKT14a}. Since investigating the dynamics of thermal instability is beyond the scope of this study, we use the opacity of solid grains, $\kappa=0.0528445 T_{\rm c}^{0.738}$, even if temperature is higher than the evaporation temperature of solid grains. 
This treatment cannot trace the variability of mass accretion rate in a short period (typically $10^2$--$10^3$ years). 
However, we consider that the time-averaged mass accretion rate is correctly calculated because the averaged mass accretion rate is determined at the outer region of the disk that is thermally stable. Owing to this treatment of thermal process, we can avoid shortening the timestep. This enables us to investigate a wide parameter range.

 \subsection{Infall from a pre-stellar core}

To obtain $\dot \Sigma_{\rm inf}$, we use the model of disk formation based on \citet{TIM13a}. We set a critical BE density profile with central density $\rho_{\rm env,0}$ and sound speed $c_{\rm s,env}$ as an initial state of the pre-stellar core. The radius of this core is $R_{\rm core}=6.46 c_{\rm s,env}/\sqrt{4\pi G \rho_{\rm env,0}}$, which is so called critical Bonnor-Ebert radius. Then, we increase the density as a factor of $f_{\rm grav}$ to make the gravity stronger than the pressure gradient. We assume that the core rigidly rotates with angular velocity $\Omega_0$. 

Because of the density enhancement, the core starts gravitational collapse, and forms the disk and star system at the center of the pre-stellar core. We divide the pre-stellar core into spherical shells, and suppose that each shell of initial radius $R$ simultaneously accretes onto the disk and star. The time when the shell accretes is 
 \begin{equation}
  t_{\rm inf}(R)=\sqrt{R^3\over GM_{\rm core}(R)}\int_0^1 {dR'\over \sqrt{\ln(R')/f_{\rm grav}+1/'R -1}},
 \end{equation}
where $M_{\rm core}(R)=\int_0^R 4\pi R'^2 \rho_{\rm env}(R') dR'$ is the enclosed mass inside the shell of radius $R$. The integral represents the delay of gravitational collapse due to the pressure gradient. The total mass accretion rate onto the disk and star is $\dot M_{\rm inf}=4\pi R^2 \rho_{\rm env}(R) dR/dt_{\rm inf}$. We assume that the angular momentum of each fluid element conserves during the collapse, and the fluid element accretes at the radius where the gravitational force balances the centrifugal force.
Then, the mass accretion rate at each radius is described as 
 \begin{equation}
  \dot \Sigma_{\rm inf} = {\dot M_{\rm inf}\over 4\pi \Omega_0 R^2 r}\delf{j}{r}\left(1-{j\over \Omega_0 R^2}\right)^{-1/2}.
 \end{equation}
The gas infall from the pre-stellar core to the disk ends at $t=t_{\rm fall}\equiv t_{\rm inf}(R_{\rm core})$. For $t>t_{\rm fall}$, $\dot M_{\rm inf}=0$ and $\dot \Sigma_{\rm inf}=0$. See \citet{TIM13a,TTM16a} for more detail of the model of disk formation.
This model reproduces a disk structure obtained by a three-dimensional hydrodynamical simulation of \citet{MIM10a}.

As an initial condition of star and disk, we put a protostar and a disk of $5\times 10^{-3} M_{\odot}$. We use the self-similar solution of \citet{LP74a} as an initial profile of surface density. Note that the initial profile is unimportant because the disk immediately forgets the initial condition due to the mass accretion from the pre-stellar core and viscous evolution. The time $t=0$ is the instant at which the core begins to collapse.

\subsection{Photoevaporation}\label{sec:photoeva}

We consider photoevaporation by X-rays from the central star. 
Although the extreme ultra-violet (EUV) photons have also ability to generate disk winds, they cannot reach the dense region of the disk due to their large cross-section. Thus, the mass loss rate by EUV photons can be neglected compared to that by X-rays \citep[e.g.,][]{OEC10a,Owe12a}.

We use the total mass loss rate and mass loss profile provided by \citet{Owe12a}. In their model, the total mass loss rate, $\dot M_{\rm X}$, and mass loss profile, $\dot \Sigma_{\rm X}$, are
\begin{equation}
 \dot M_{\rm X} = 6.25\times 10^{-9}\left({M_*\over M_\odot}\right)^{-0.068}\left({L_{\rm X}\over 10^{30}\rm erg~s^{-1}}\right)^{1.14}\rm M_\odot~ yr^{-1},
\end{equation}
\begin{equation}
 \dot \Sigma_{\rm X}(x>0.7) = \dot \Sigma_0 \xi(x) \exp\left[-\left({x\over 100}\right)^{10}\right] ,
\end{equation}
\begin{equation}
x(r,~M_*)=0.85\left({r \over \rm au}\right)\left({M_*\over M_\odot}\right)^{-1},
\end{equation}
where $L_{\rm X}$ is the X-ray luminosity from the central star (see the next subsection for prescription of $L_{\rm X}$) and $\dot \Sigma_0$ is the normalization factor determined by the condition $\dot M_{\rm X} = \int_0^\infty 2\pi r \dot \Sigma_{\rm X} dr $. 
We set $\dot \Sigma_{\rm X}=0$ for $x<0.7$.
In the late phase, this X-ray driven winds creates an inner hole in the disk. We evaluate the inner hole radius, $r_{\rm hole}$, such that $\tau_{\rm X}=1$, where $ \tau_{\rm X} = \int_{r_{\rm in}}^{r_{\rm hole}} \sigma_{\rm X} \rho_0 dr/(\mu m_p) $ is the optical depth for X-rays from the central star ($\sigma_{\rm X}\sim 10^{-22}$ cm$^2$ is the  X-ray cross-section for a particle of gas). The existence of an inner hole modifies the mass loss rate and profile. If $x(r_{\rm hole},M_*) > 0.7$ is satisfied, we use the mass loss rate and profile with an inner hole written as
\begin{equation}
 \dot M_{\rm X} = 4.8\times 10^{-9}\left({M_*\over M_\odot}\right)^{-0.148}\left({L_{\rm X}\over 10^{30}\rm erg~s^{-1}}\right)^{1.14}\rm M_\odot~ yr^{-1},
\end{equation}
\begin{equation}
 \dot \Sigma_{\rm X}(y\ge0) = \dot \Sigma_0 \zeta(y) \exp\left[-\left({y\over 57}\right)^{10}\right],
\end{equation}
\begin{equation}
y(r-r_{\rm hole},~M_*)=0.95\left({r-r_{\rm hole} \over \rm au}\right)\left({M_*\over M_\odot}\right)^{-1}.
\end{equation}
See Appendix B of \citet{Owe12a} for the detailed formulae of $\xi(x)$ and $\zeta(y)$.

After the hole formation, there remains a little amount of gas inside the hole. This gas can absorb the X-rays, so that the disk wind should blow even for $y<0$. Since absorption rate of the X-rays is proportional to the density, we assume the following formula for $y<0$,
\begin{equation}
 \dot \Sigma_{\rm X}(y<0) = \dot \Sigma_{\rm X}(y=0) \min\left({\Sigma(y)\over \Sigma(y=0)},0.1 \right).
\end{equation}
We set $\dot \Sigma_{\rm X}=0$ for $x<0.7$.

It is likely that the photoevaporation is suppressed during the infall phase. Thus, for $t<t_{\rm fall}$, we set $\dot M_{\rm X}=0$.

  \begin{figure}
   \includegraphics[width=\linewidth]{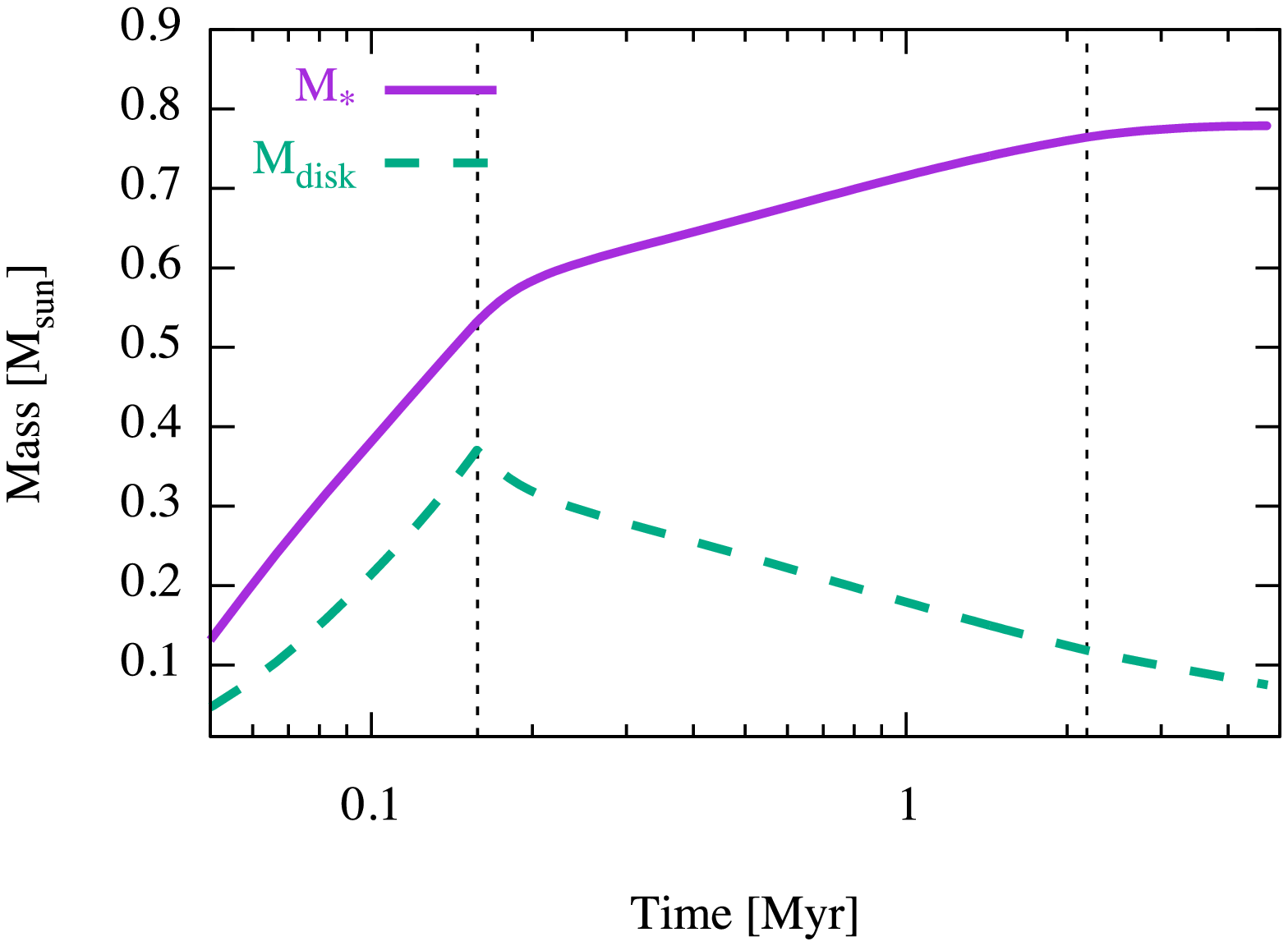}
   \includegraphics[width=\linewidth]{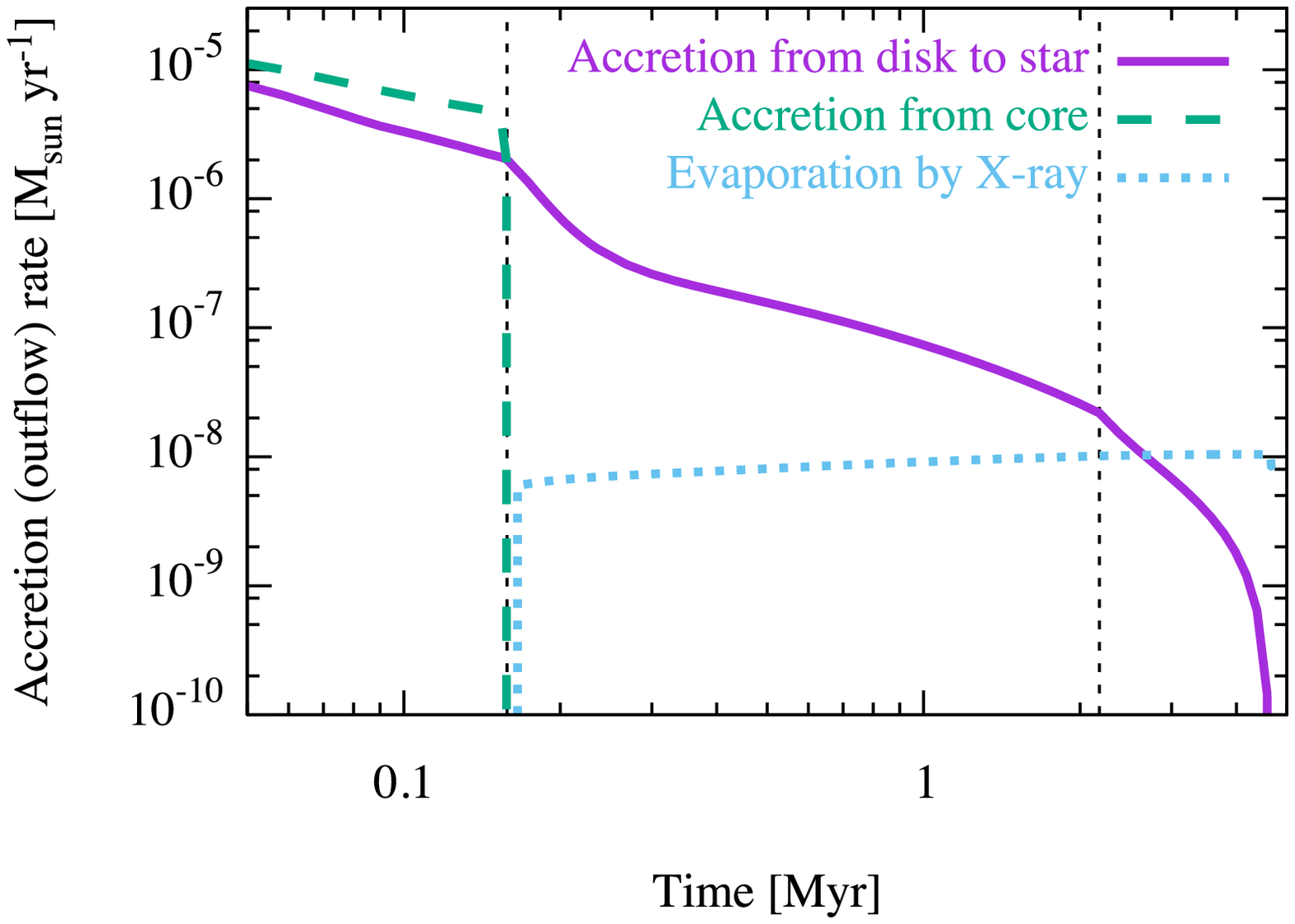}
   \includegraphics[width=\linewidth]{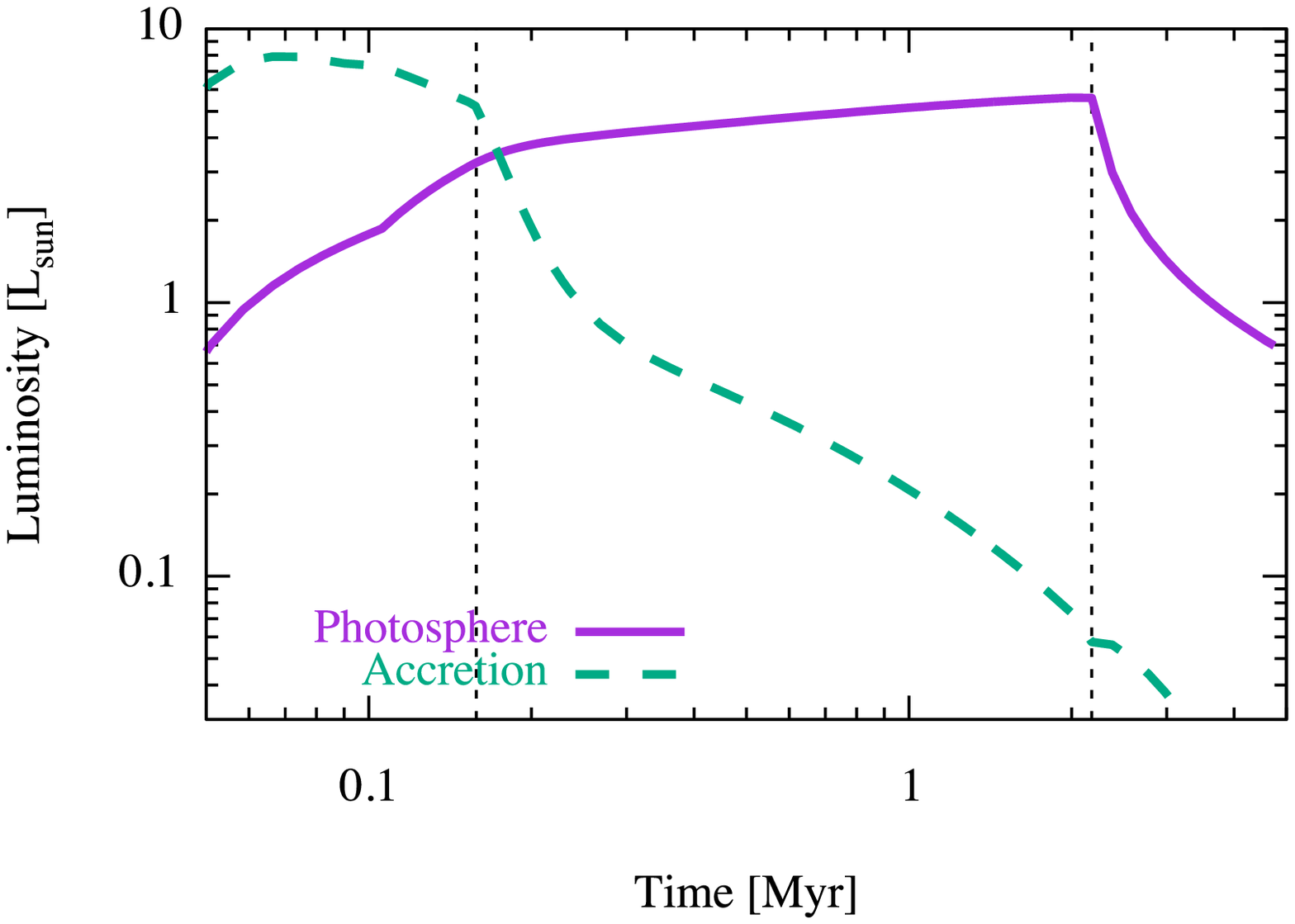}
   \caption{The time evolution of physical quantities for run A2. The top panel shows  $M_*$ (solid) and $M_{\rm disk}$ (dashed). The middle panel depicts mass accretion rate from disk to star (solid), $\dot M_{\rm inf}$ (dashed), and $\dot M_X$ (thick-dotted). The bottom panel represents $L_{*,\rm ph}$ (solid) and $L_{\rm accrt}$ (dashed). The left and right thin-vertical dotted lines show $t_{\rm fall}$ and $t_{\rm pms}$, respectively. 
   \label{fig:timeevolution}}
  \end{figure}


\subsection{Evolution of central stars}

We need to take the stellar evolution into account because its luminosity is important for the disk temperature.
When the protostellar mass is increasing, we assume that the protostar evolves along the birthline provided by \citet{SP05a}, 
which corresponds to the evolutionary track of an accreting star. Although \citet{BCG09a} has pointed out that entropy brought to accreting stars through accretion can modify the evolutionary tracks of the accreting stars significantly, we use a standard evolutionary model for simplicity.
When the growth of stellar mass is completed, we use the evolutionary tracks for pre-main sequence stars of \citet{SDF00a}. We judge the end of growth of the stellar mass by the condition that the disk evaporation timescale becomes sufficiently shorter than the growth timescale of stellar mass, $t_{\rm grow} \ge 3t_{\rm evap}$, where $t_{\rm evap}=M_{\rm disk}/\dot M_{\rm X}$ and $t_{\rm grow}= M_*/\dot M_*$ ($\dot M_*$ is the mass accretion rate onto the central star). We define the instant when this condition is satisfied as the starting time of pre-main sequence phase, $t_{\rm pms}$. We have confirmed that the growth of the stellar mass is less than $4\%$ after $t_{\rm pms}$. 
We obtain the photospheric stellar luminosity, $L_{*,\rm ph}$, and the stellar  radius, $R_*$, as a function of the stellar mass and time from the data sets of evolutionary tracks by using the linear interpolation. We also calculate the accretion luminosity, $L_{\rm accrt} = |GM_*\dot M_*/(2R_*)|$, and use the sum as the stellar luminosity $L_*=L_{\rm accrt}+L_{*,\rm ph}$.

We use the X-ray luminosity suggested by \cite{PKF05a}: $L_{\rm X}=L_{\rm X,0}\left({M_*/M_\odot}\right)^{1.44}$, where $L_{\rm X,0}$ is a parameter in the present paper. The mean value and standard deviation of the parameter $L_{\rm X,0}$ are set to be $2.3\times 10^{30} ~\rm erg~s^{-1}$ and $0.65$ dex, respectively.

\subsection{Disk dispersal condition}

The disk fraction of each cluster is usually determined by observations of near-infrared \citep[NIR, e.g.][]{HLL01a}, and the mean disk lifetime is derived by the integration of disk fractions of clusters. Thus,
we use the disk dispersal condition such that the optical depth in the NIR-emitting region becomes lower than unity. We define the NIR-emitting region as a hotter region than 300 K because NIR ranges from 1 to 8 $\rm\mu$m. The disk dispersal time in NIR range, $t_{\rm NIR}$, is defined as the instant when $\tau<1$ is satisfied in the region where $T_{\rm c}>$300 K. Observationally, the age of young clusters is determined by using the evolutionary tracks of pre-main sequence stars. The age is not defined as the time after the collapse of a pre-stellar core, but defined as the time after the pre-main sequence phase starts. Thus, we estimate the disk lifetime of NIR to be $t_{\rm life}=t_{\rm NIR}-t_{\rm pms}$.

\section{Results} \label{sec:results}

\subsection{Typical evolution of a protoplanetary disk}\label{sec:typical}

\begin{table}
\begin{center}
\caption{Parameter sets for each runs. \label{tab:param}}
\begin{tabular}{|c|ccc|}
\hline
 Runs & $n_{\rm env,0}$ & $\Omega_0$ & $L_{\rm X,0}$  \\
 & [10$^6$ cm] & [km s$^{-1}$ pc$^{-1}$] & [$L_{\rm X,m}$]\\
\hline
\hline
A1  & 2.0 & 0.5 &   1.0 \\
A2 (reference) & 2.0& 1.5& 1.0 \\
A3  & 2.0& 2.5& 1.0 \\
\hline
B1  & 0.5 & 1.5 & 1.0 \\
B2  & 8.0&1.5& 1.0 \\
\hline
C1  &  2.0 & 1.5 &  0.2 \\
C2  & 2.0 & 1.5 & 5.0 \\
\hline
\end{tabular}
\end{center}
\end{table}

We calculate the evolution of protoplanetary disks, taking account of formation and evaporation. We set computational region from 0.01 to 10$^4$ au. We equally divide the region in logarithmic space with 100 meshes, and solve Equation (\ref{eq:diffusion}) explicitly using the forward-time centered-space diferencing scheme \citep{pre+92}. 
We fix $f_{\rm grav}=1.4$, $c_{\rm s,env}=1.9\times10^4$ cm s$^{-1}$, $T_{\rm amb}=10$ K, and $\alpha_{\rm MRI}=0.01$. We calculate several values of $\Omega_0$, $n_{\rm env,0}$, and $L_{\rm X,0}$ as tabulated in Table \ref{tab:param}.

  \begin{figure}
   \includegraphics[width=\linewidth]{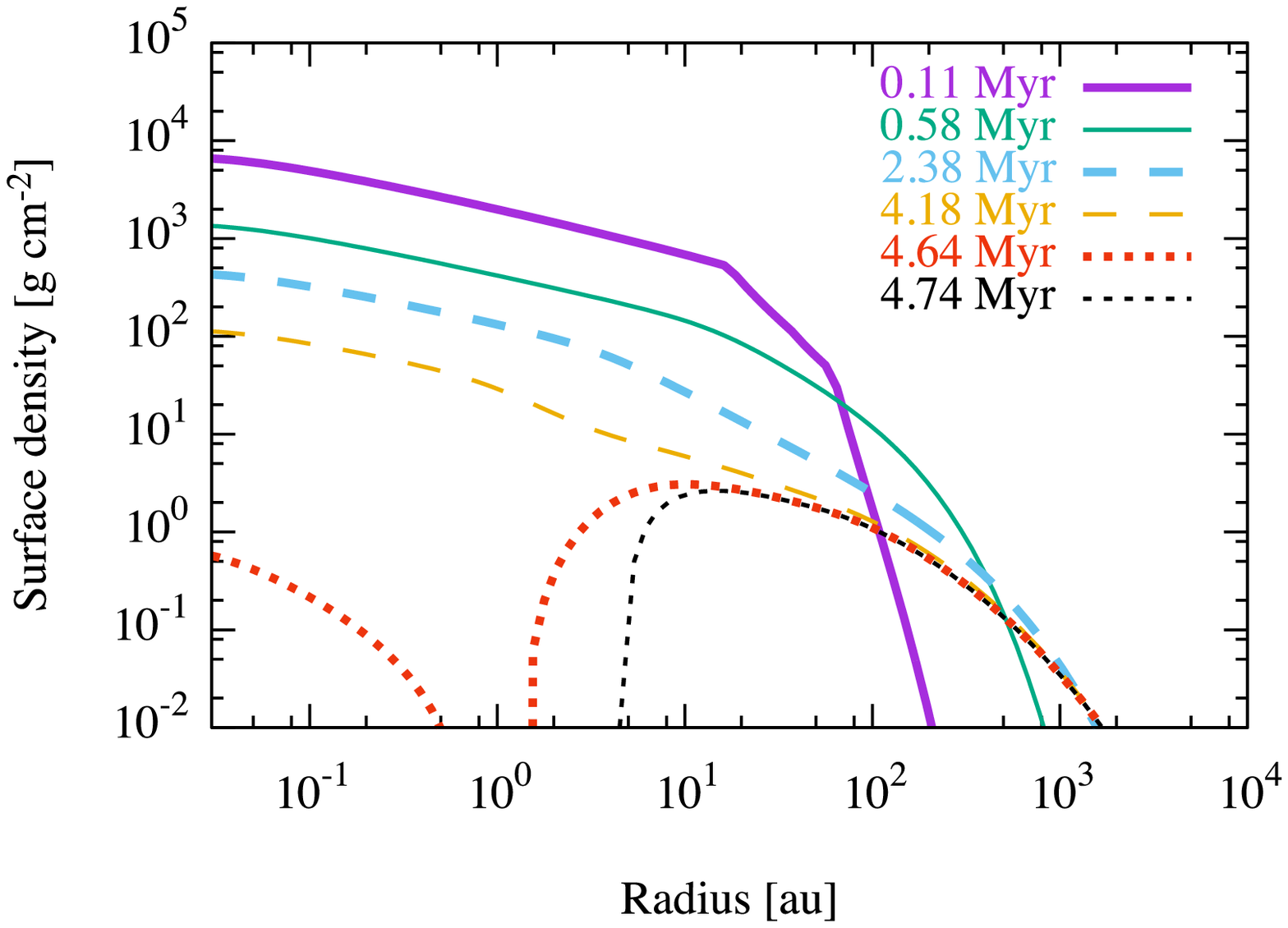}
   \includegraphics[width=\linewidth]{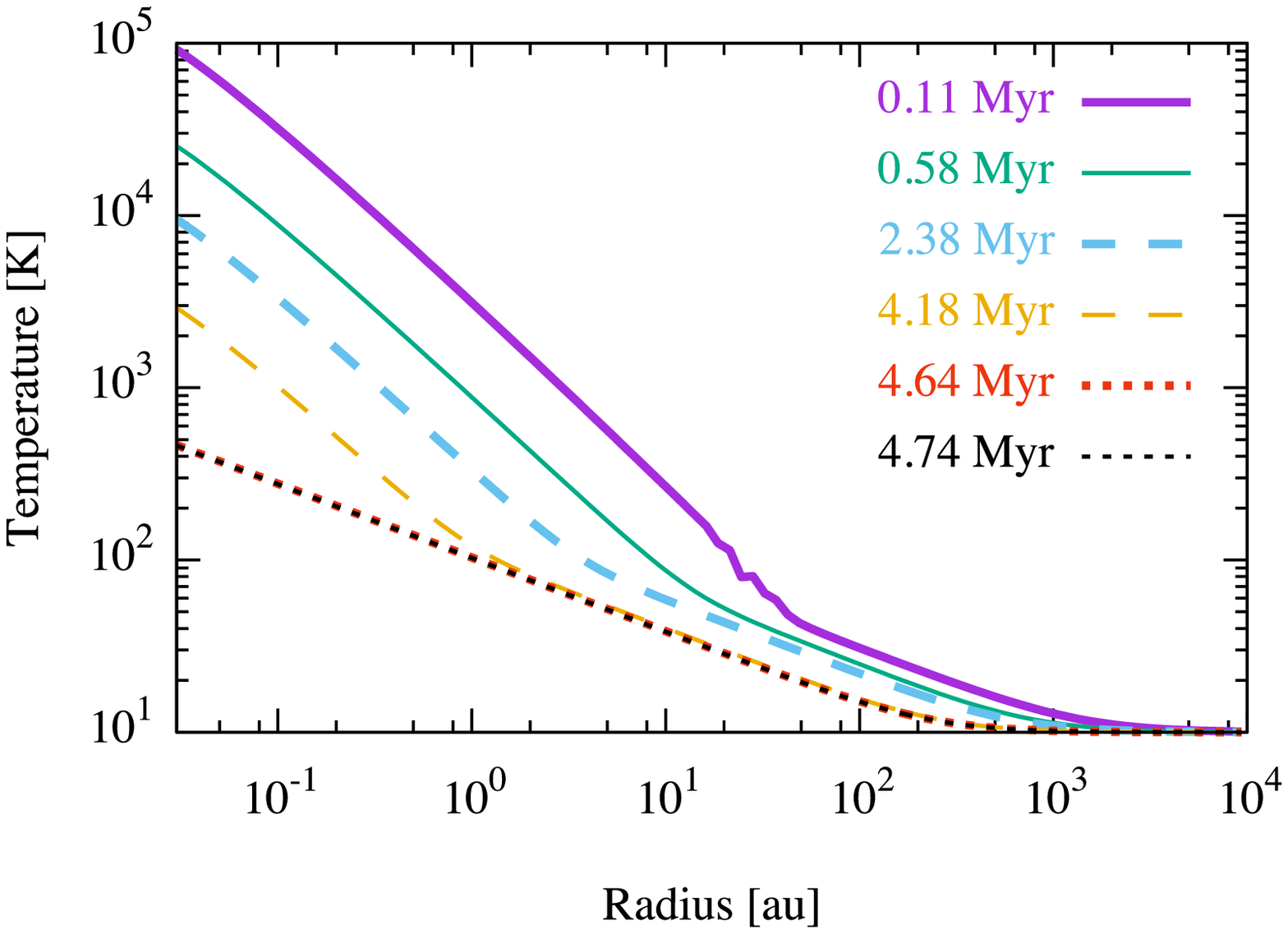}
   \caption{The time evolution of radial profile of surface density (upper) and temperature (lower) for run A2. 
   \label{fig:sigtemp}}
  \end{figure}

Figure \ref{fig:timeevolution} shows the time evolution of physical quantities for run A2. There are three phases: infall phase $t<t_{\rm fall}$, viscous evolution phase $t_{\rm fall}<t<t_{\rm pms}$, and pre-main sequence phase $t>t_{\rm pms}$. In the infall phase, the mass accretion rate from cloud core is as high as $\simeq 5\times10^{-6}~M_{\odot}~\rm yr^{-1}$, and both the protostar mass and the disk mass increase. The gravitational torque dominates over the MRI turbulence for the mechanism of angular momentum transport during this  phase. After infall finishes, the disk mass decreases while the protostar continues to grow up, owing to the high mass accretion rate from the disk onto the protostar. The MRI turbulence mainly drives this mass accretion, and the typical mass accretion rate is about $10^{-8}$--$10^{-7}~M_{\odot}~\rm yr^{-1}$, which is consistent with observation of T-Tauri stars.
After a few Myr, the disk becomes sufficiently lighter than the central star, and the pre-main sequence phase starts because the condition $t_{\rm evap} < 3 t_{\rm grow} $ is satisfied. For $t>t_{\rm pms}$, the central star becomes fainter because of Kelvin-Helmholtz contraction. The mass accretion rate onto the central star continues to decrease, while the photoevaporation is efficient. Thus, the disk becomes lighter and lighter, and finally, the disk disperses. 

The upper panel of Figure \ref{fig:sigtemp} shows the radial profile of surface density with some time snapshots for run A2. In the early phase of disk evolution, the disk spreads outward due to the diffusion nature of viscous accretion disks \citep[e.g.][]{LP74a,pri81}. The cutoff radius $r_{\rm cut}$  (kind of disk radius; see Appendix for the definition of $r_{\rm cut}$) increases with time, and is larger than $r_{\rm cent}$  at $t=t_{\rm pms}$ as shown in Table \ref{tab:results}, where $r_{\rm cent}=R_{\rm env}^4 \Omega_0^2/(G M_{\rm env})$ is the centrifugal radius of outermost region of the pre-stellar core. Although the photoevaporation process has little effect on disk evolution during the infall and viscous evolution phases, it determines the disk structure in pre-main sequence phase. Since the mass loss profile by photoevaporation has a peak, the gap appears around 1 au \citep[e.g.][]{OEC10a,BHZ13a}. After the gap formation, the inner disk that is detached from the outer disk accretes onto the central star in its viscous timescale ($\sim 10^5$ years), which is much shorter than other timescales, such as the disk evaporation timescale ($\sim 10^7$--$10^8$ years). Thus, the gap becomes an inner hole. Then, the mass loss profile changes to that with an inner hole (see Section \ref{sec:photoeva}). This profile has a sharp peak at the inner edge of the disk. Therefore, the disk is dissipated from the inside to outside \citep{CGS01a,OEC10a}.

The bottom panel of Figure \ref{fig:sigtemp} shows the radial profile of temperature for run A2. During the infall phase, viscous heating dominates over the other heating mechanisms inside the cutoff radius. 
The heating by the infalling material is always sub-dominant. After the infall finishes, the irradiation heating determines the temperature for outer region of the disk (e.g. $r \gtrsim 10$ au at $t=0.58$ Myr). As surface density decreases, the viscous heating rate is gradually decreasing, and the radius at which  $T_{\rm irr} = T_{\rm vis}$ moves inward. After $t_{\rm pms}$, 
the stellar evolution results in the decrease of the irradiation temperature with time. Note that the duration of the Hayashi phase of solar-mass stars is about 10$^7$ years.

  \begin{figure}
   \includegraphics[width=\linewidth]{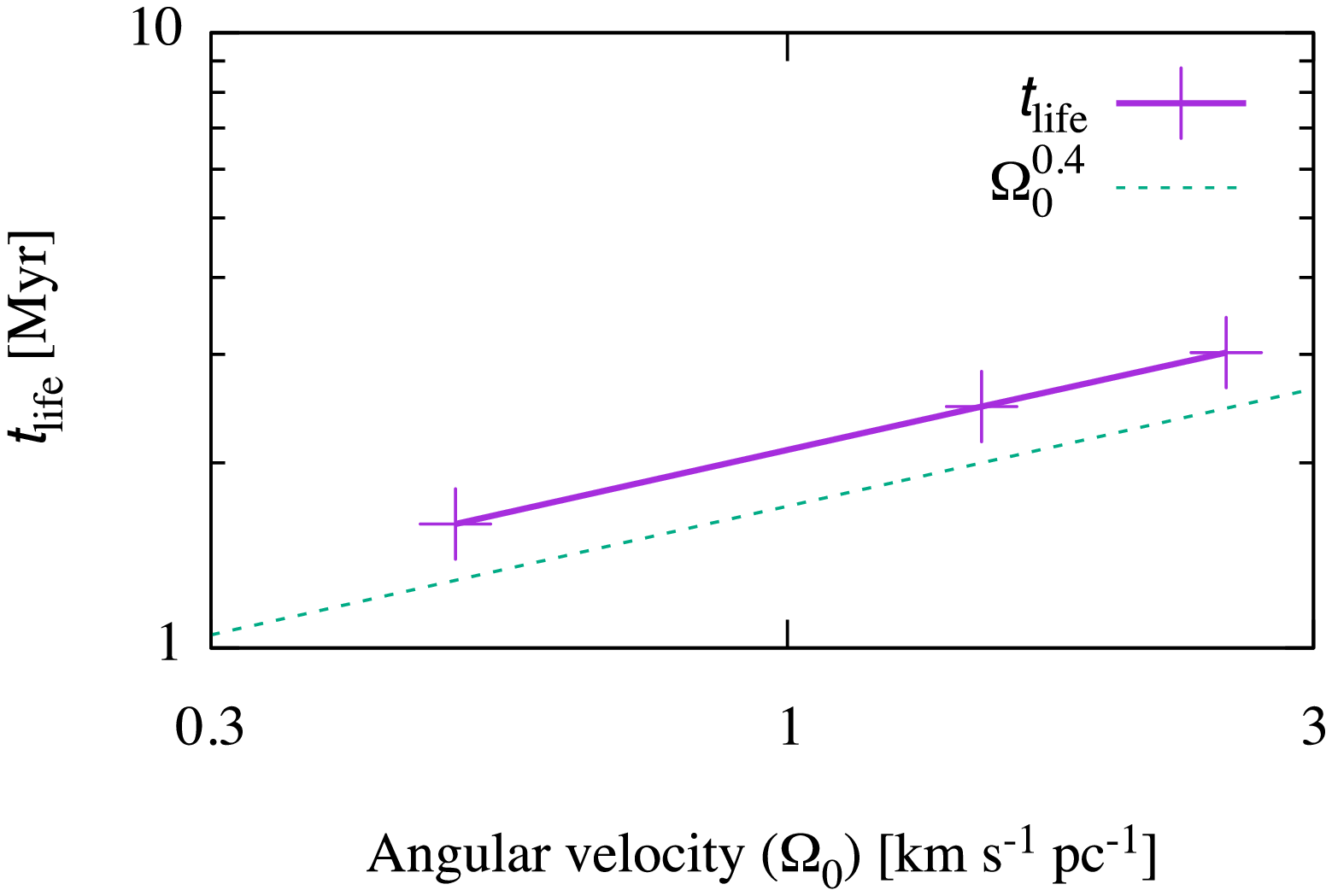}
   \includegraphics[width=\linewidth]{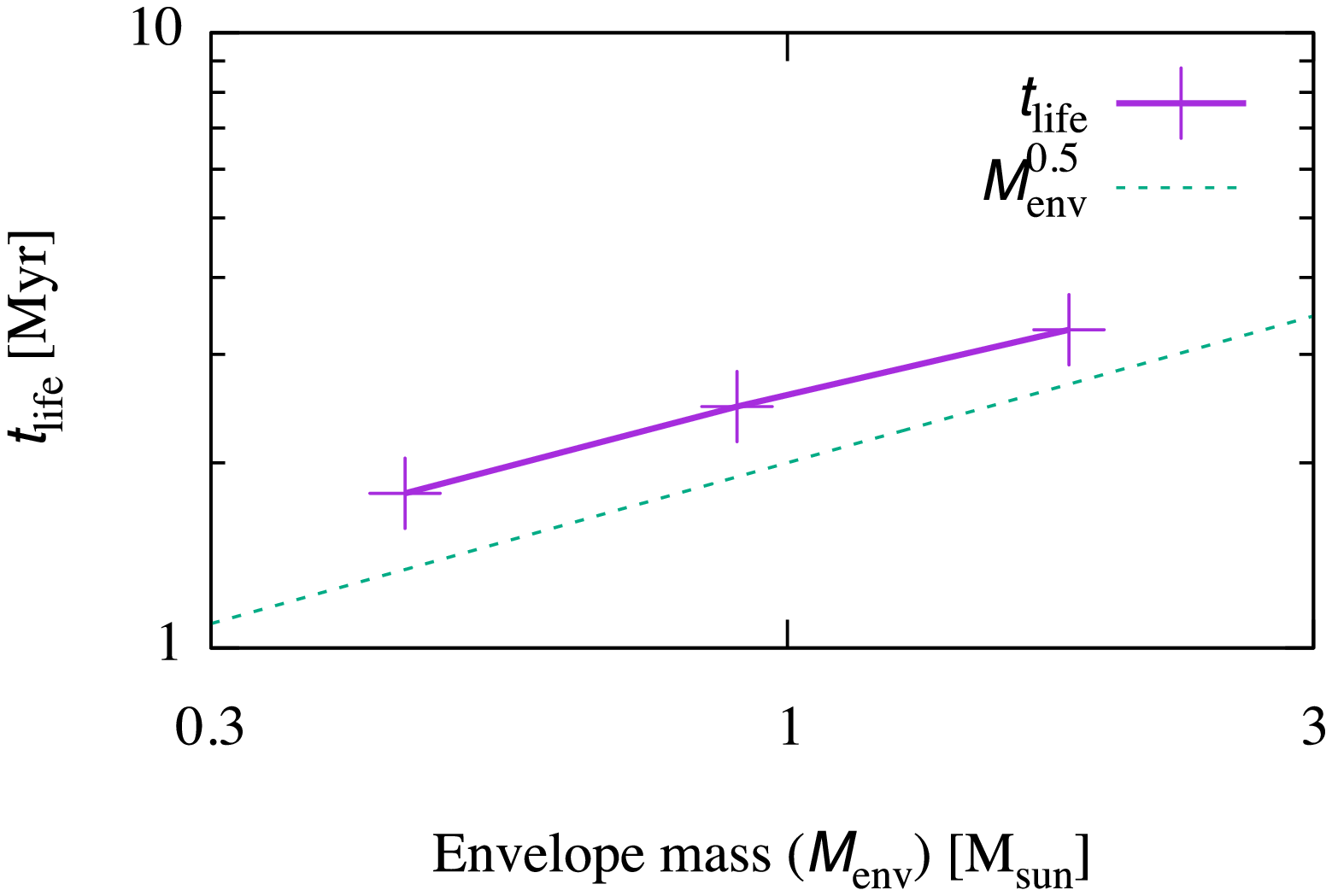}
   \includegraphics[width=\linewidth]{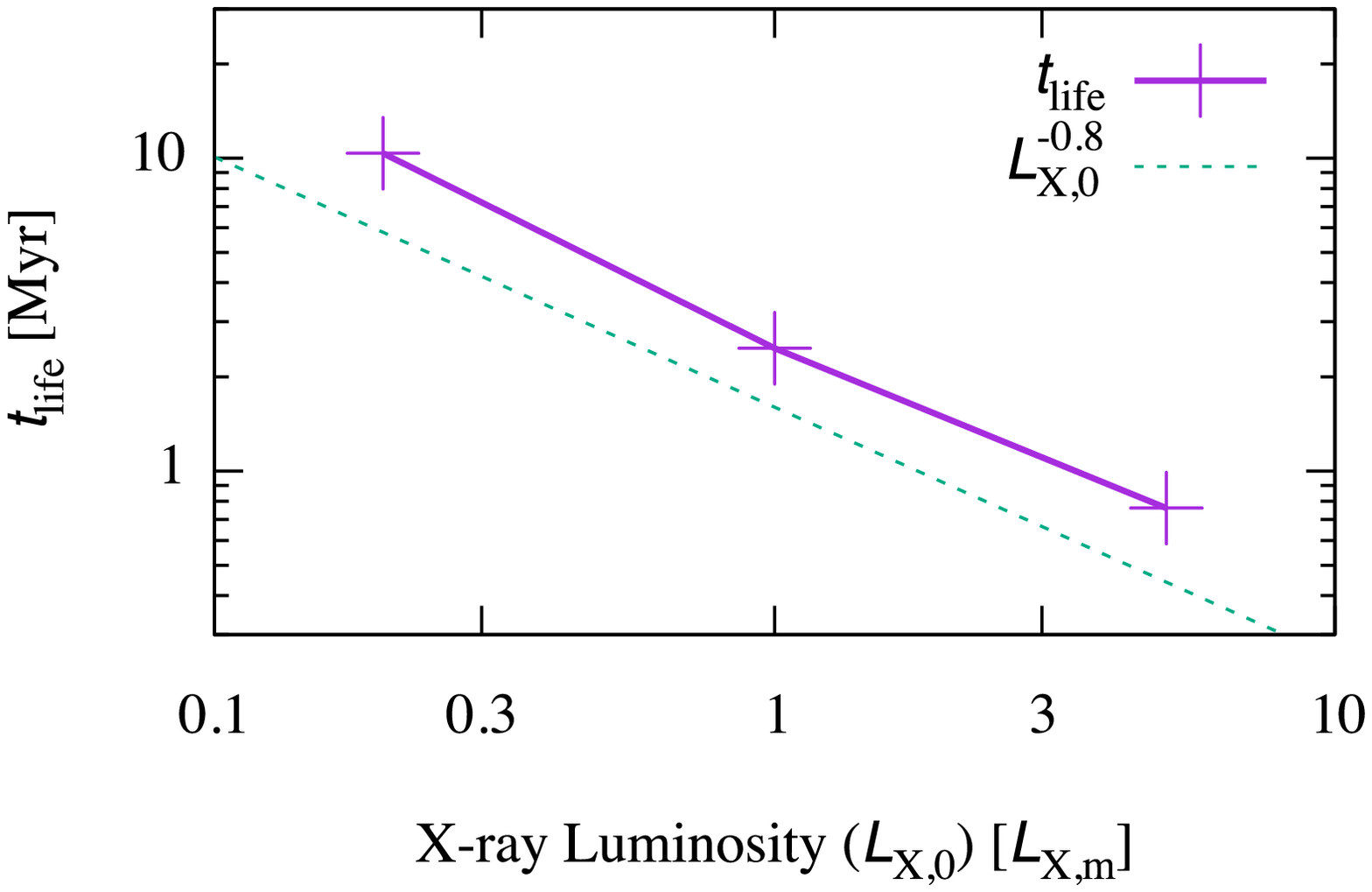}
   \caption{The disk lifetime $t_{\rm life}$ as a function of $\Omega_0$ (upper), $M_{\rm env}$ (middle), and $L_{X,0}$ (lower). The dotted lines show the dependence of disk lifetime on these parameters. 
   \label{fig:tlife}}
  \end{figure}

\begin{table*}
\begin{center}
\caption{Resultant quantities for each runs. For $M_*$ and $M_{\rm disk}$, we write their values at both $t=t_{\rm fall}$ and $t=t_{\rm pms}$.  For $r_{\rm cut}$, the values at $t=t_{\rm pms}$ is described. \label{tab:results}}
\begin{tabular}{|c|ccccccccc|}
\hline
 Runs  & $t_{\rm fall}$ & $t_{\rm pms}$  & $t_{\rm life}$ 
 & $M_{\rm core}$
 & $M_*$
 & $M_{\rm disk}$
 & $R_{\rm core}$ 
 & $r_{\rm cent}$
 & $r_{\rm cut}$ \\
 & [Myr] &[Myr] &[Myr] & [M$_{\odot}$] &[M$_{\odot}$] &[M$_{\odot}$] & [10$^4$ au] & [10$^2$ au] & [10$^2$ au]\\
\hline
\hline
A1 &  0.16 &  0.73 & 1.6 & 0.90 & 0.73/0.82 & 0.17/0.071 & 6.5 & 0.14 & 1.9  \\
A2 (reference) & 0.16 & 2.2 & 2.5& 0.90 & 0.53/0.76 &  0.37/0.12 & 6.5 & 1.3 & 6.8 \\
A3  & 0.16 & 3.5 & 3.0& 0.90 & 0.44/0.73 &  0.47/0.15 & 6.5 & 3.4 & 12\\
\hline
B1  & 0.32 & 5.1 & 3.3 & 1.8 & 0.80/1.3 & 1.0/0.44 & 13 & 8.9 & 15\\
B2  & 0.079 &  0.88 & 1.8 & 0.45 & 0.34/0.42 & 0.11/0.032 & 3.2 & 0.16 & 3.0\\
\hline
C1  & 0.16 & 5.3 & 10 & 0.90 &  0.53/0.82 & 0.37/0.074 &   6.5 & 1.3 & 15 \\
C2  & 0.16 & 0.81 & 0.76 & 0.90 & 0.53/0.69 & 0.37/0.18 & 6.5 & 1.3 & 3.0 \\
\hline
\end{tabular}
\end{center}
\end{table*}

We tabulate the resultant physical quantities in Table \ref{tab:results}. All the calculations qualitatively have the same evolutionary characteristics described above. We find that the timescales are longer for lower $n_{\rm env,0}$, higher $\Omega_0$, and lower $L_{\rm X,0}$. The cores with higher $\Omega_0$ have higher angular momenta. This makes the disk radius larger, which leads to the longer disk evolution timescales $t_{\rm pms}$ and $t_{\rm life}$. For lower $n_{\rm env,0}$, $t_{\rm fall}$ is longer, since $t_{\rm fall}$ is proportional to the free-fall time. The cores with lower $n_{\rm env,0}$ have higher angular momenta owing to the larger initial disk radii, so that they have the longer  $t_{\rm pms}$ and $t_{\rm life}$. The photoevaporation rate is higher for higher $L_{\rm X,0}$, which results in the shorter  $t_{\rm pms}$ and $t_{\rm life}$. We also find that the disk lifetime has a power-law dependence on parameters, $t_{\rm life}\propto \Omega_0^{0.4}$, $\propto M_{\rm env}^{0.5}$, and $\propto L_{\rm X,0}^{-0.8}$ as illustrated in Figure \ref{fig:tlife}.
Thus, the stellar X-ray luminosity has the strongest impact on the disk lifetime.

For runs with higher $\Omega_0$, lower $n_{\rm env,0}$, and lower $L_{\rm X,0}$, the duration of the accretion phase ($t_{\rm pms}-t_{\rm fall}$) seems to be longer ($\sim$ 3--5 Myr) than expected \citep{EDJ09a,MAMK11a}.
If this long accretion phase is real under a system with a high $\Omega_0$, the age spread of pre-main sequence stars in clusters \citep[e.g.][]{Hil09a} can be solved by the scatter of angular momenta of pre-stellar cores. It is also expected that the longer accretion phase arises from our treatment of stellar evolution, for which the pre-main sequence phase starts when the growth of the protostar stops. It is expected that the pre-main sequence phase starts when the timescale of Kelvin-Helmholtz contraction becomes shorter than the growth time of the central star. 
This condition is approximately written as $L_{*,\rm ph} > L_{\rm accrt}$, which is satisfied for $t < t_{\rm pms}$ (see the bottom panel of Fig. \ref{fig:timeevolution}). In order to obtain a solid conclusion, we should simultaneously solve the evolution of the central star with time-dependent mass accretion, which is beyond the scope of this paper.

\subsection{Disk fraction}\label{sec:diskfraction}

\begin{figure}
   \includegraphics[width=\linewidth]{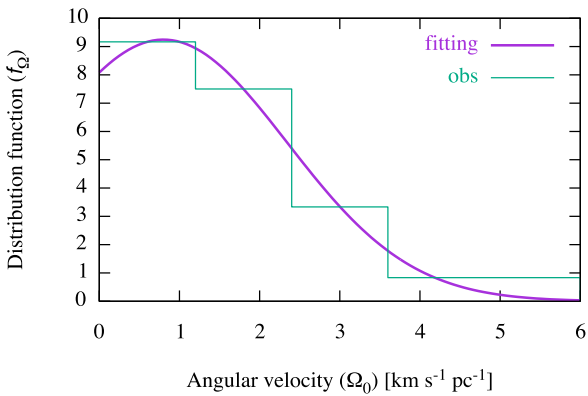}
   \caption{The distribution function for $\Omega_0$. The thin line shows the observational data, and the thick line shows the fitting result.
   \label{fig:distribution}}
  \end{figure}

For comparison of our results to observations, we derive disk fraction as a function of stellar age. To obtain the disk fraction, we consider the distribution functions for $\Omega_0,~M_{\rm env}$, and $L_{\rm X,0}$. From observations of pre-stellar cores, we can obtain the distribution function of  $\Omega_0$ and the core mass function (CMF; the distribution function of $M_{\rm core}$). \citet{CBM02a} observed velocity gradients of 26 pre-stellar cores. We use their data to obtain the distribution function of $\Omega_0$. We find that the distribution function is well-fitted by the normal function,
\begin{equation}
 f_\Omega={dN\over d\Omega_0}= C_1 \exp\left[{\left(\Omega_0-\Omega_{\rm m}\right)^2\over 2 \sigma_\Omega^2}\right],
\end{equation}
with $\Omega_{\rm m}=0.8$ km s$^{-1}$ ps$^{-1}$ and $\sigma_\Omega=1.5$ km s$^{-1}$ ps$^{-1}$. We plot the observational data and fitting function of the distribution function in Figure \ref{fig:distribution}, in which we see that the fitting formula matches the observation well. For CMF, we use CMF of Aquila given in \citet{KAM10a}, which is represented as a log-normal function,
\begin{equation}
 f_M= {dN\over d\log M_{\rm core}} = C_2 \exp\left[{\left(\log M_{\rm core}-\log M_{\rm m}\right)^2\over 2 \sigma_M^2}\right],
\end{equation}
where $M_{\rm m}=0.9\rm M_{\odot}$ and $\sigma_M=0.3$ are the peak and dispersion of the log-normal function, respectively. In addition to the distributions of pre-stellar cores, we consider the distribution of $L_{\rm X,0}$ from X-ray observations of YSOs. Assuming the distribution function for $L_{\rm X,0}$ is the log-normal, we can write it as 
\begin{equation}
 f_L=  {dN\over d\log L_{\rm X,0}} = C_3 \exp\left[{\left(\log L_{\rm X,0}-\log L_{\rm X,m}\right)^2\over 2 \sigma_L^2}\right],
\end{equation}
where $L_{\rm X,m}=2.3\times10^{30}$ erg s$^{-1}$ and $\sigma_L=0.65$ \citep{PKF05a}.

\begin{figure}
   \includegraphics[width=\linewidth]{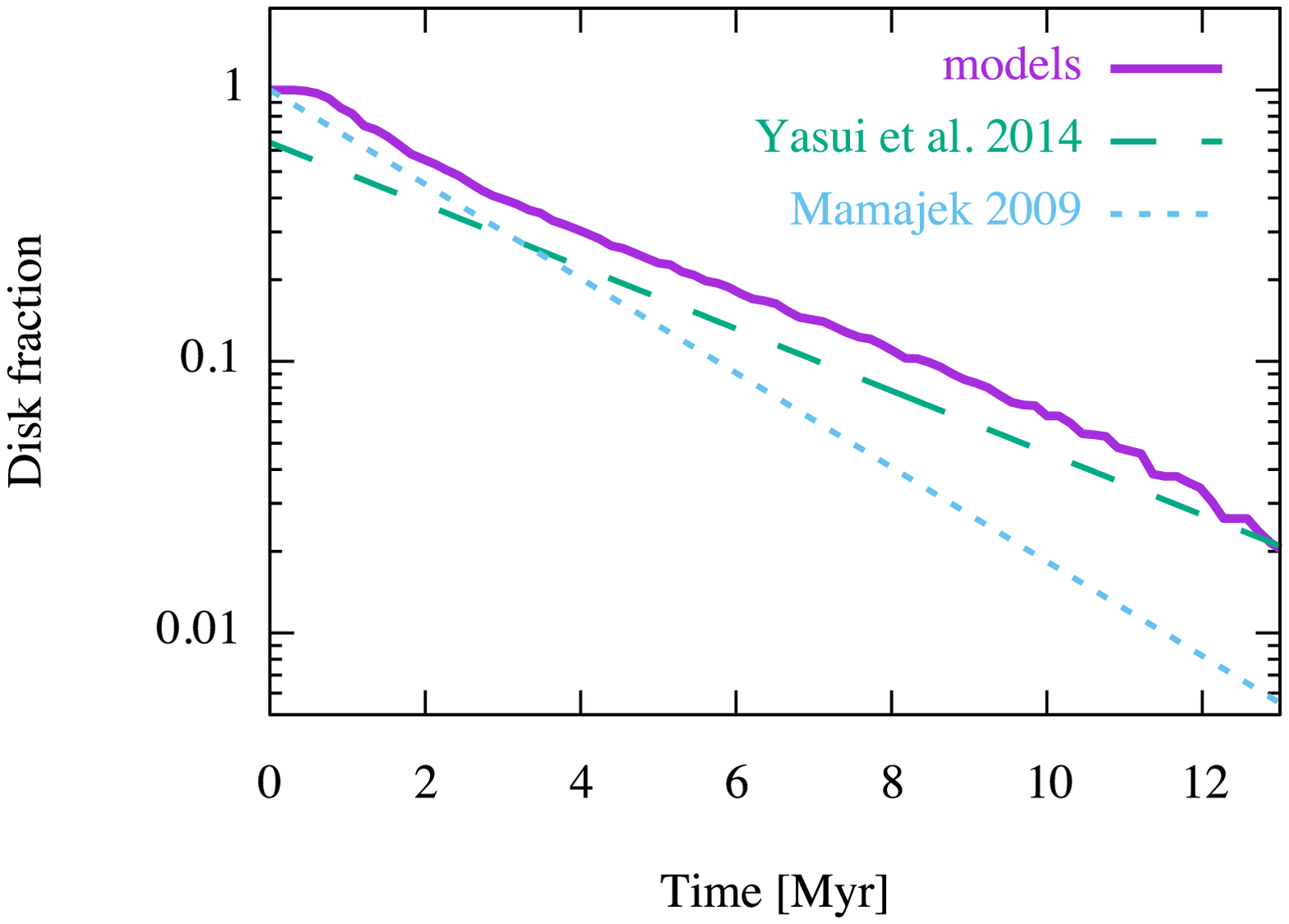}
   \includegraphics[width=\linewidth]{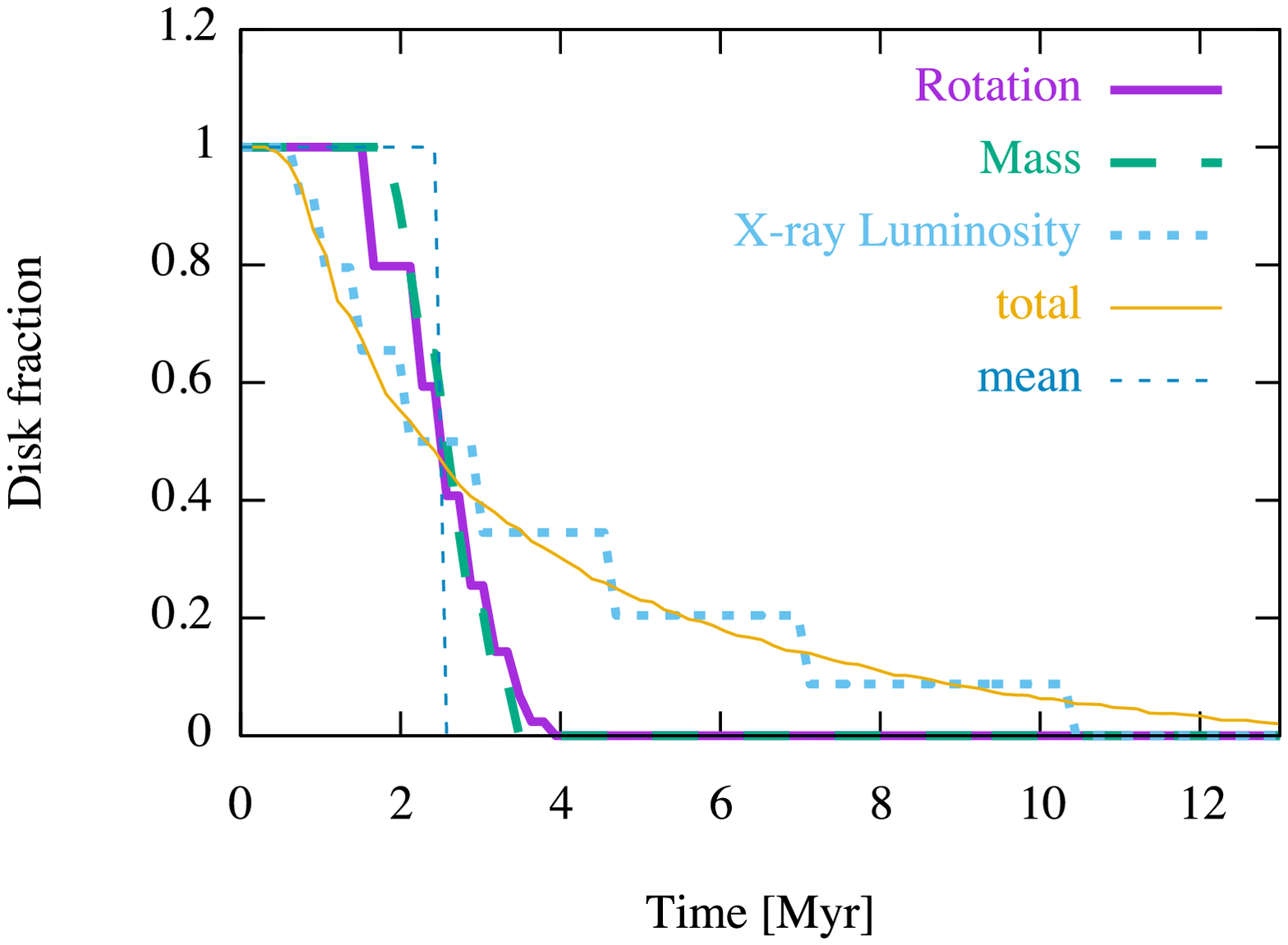}
   \caption{Calculated disk fraction as a function of time. The upper panel shows comparison of our model to observational estimates. The our model (solid line) is roughly consistent with the observational estimate by \citet{YKT14a} (dashed line). The dotted line is another observational estimate \citep{Mama11a}. 
   The lower panel shows the effect of each parameter on broadening the lifetime of disks. The effects of rotation (thick-solid) and envelope mass (thick-dashed) is sub-dominant, while the X-ray luminosity (thick-dotted) mainly determines the disk lifetime. The model prediction shown in upper panel (thin-solid) and mean lifetime (thin-dotted) are also shown.
   \label{fig:diskfraction}}
  \end{figure}

  Assuming that these parameters have no correlation, the disk fraction at the stellar age $t'$ is represented as 
  \begin{equation}
  f_{\rm disk}(t') = {\int \int \int d\Omega_0 d\log M_{\rm core} d\log L_{\rm X,0} \Theta(t_{\rm life}-t') f_\Omega f_M f_L \over \int \int \int d\Omega_0 d\log M_{\rm core} d\log L_{\rm X,0}  f_\Omega f_M f_L}
  \end{equation}
  where $\Theta(t')$ is the Heaviside's step-function. We note that the age $t'=0$ is the instant when the pre-main sequence phase starts. We perform $8^3=512$ calculations for a parameter range of $0.5~{\rm km~s^{-1}~pc^{-1}}\le \Omega_0\le 4~{\rm km~s^{-1}~pc^{-1}}$, $0.45~M_{\odot}\le M_{\rm core}\le 1.8~M_{\odot}$, and $0.2 L_{\rm X,m} < L_{\rm X,0} < 5 L_{\rm X,m}$. We equally divide the grid of $\Omega_0$ in linear space, while the grids of $M_{\rm env}$ and $L_{\rm X,0}$ are equally divided in logarithmic space. The comparison of resultant disk fraction to that estimated from observations is shown in the upper panel of Fig. \ref{fig:diskfraction}. Our model gives the disk fraction that exponentially decreases with time $f_{\rm disk}\propto \exp(-t/t_{\rm life,m}) $, where $t_{\rm life,m}\simeq$ 3.7 Myr is the mean lifetime of disks. This lifetime is consistent with the observational estimate of 3.8 Myr obtained by \citet{YKT14a}, although there still remains uncertainty in the observational estimate (the mean lifetime is 2.5 Myr according to \citet{Mama11a}). Our model gives the slightly larger disk fraction at a given time. We discuss the reason in the next section. 

To see the parameter that mainly broaden the disk lifetime, we calculate the disk fraction varying only one parameter and fixing the other parameters as their reference values ($\Omega_0=1.5 \rm~km~s^{-1}~pc^{-1}$, $M_{\rm env}=0.9M_{\odot}$, $L_{\rm X,0}=L_{\rm X,m}$). The result is shown in the lower panel of Fig. \ref{fig:diskfraction}. Although the disk fractions are step-like because of the lack of resolution in the parameter space, we find that the X-ray luminosity mainly determines the broadening of disk fraction, whereas the angular velocity and mass of pre-stellar cores are sub-dominant. Therefore, further studies are highly encouraged to reveal the origin of the X-ray luminosities of YSOs \citep[e.g.][]{FDM03a,PKF05a,TGB07a}. This broadening of disk fraction by the dispersion in X-ray luminosities has been reported by \citet{OEC11a}. 

In this paper, we assume that the alpha parameter by MRI turbulence, $\alpha_{\rm MRI}$, is constant. For the case with low $L_{\rm X,0}$, the ionization degree in the disk decreases, and MRI can be inactivated \citep{IG99a,TS08a,FOT14a}. Since $\alpha_{\rm MRI}$ in the MRI inactive region is lower than that in MRI active region, it may be correlated with the X-ray luminosity. If this effect is included in our model, the model with weaker (stronger) $L_{\rm X,0}$ has a lower (higher) value of $\alpha_{\rm MRI}$, which results in longer (shorter) $t_{\rm life}$. This probably causes rapider decrease of the disk fraction in early phase and slower decrease in rate phase than that shown in Figure \ref{fig:diskfraction}. The impact of dead zone on the mean lifetime, $t_{\rm life,m}$, remains as a future work.


\section{Summary} \label{sec:summary}

We have calculated the evolution of protoplanetary disks, taking account of formation from pre-stellar cores and dispersal by X-ray photoevaporation. A typical pre-stellar core collapses to form the star-disk system in 0.2 Myr. After infall from the pre-stellar core finishes, the star mass increases through accretion from the disk, while the disk mass gradually decreasing. Then, a few Myr later, mass accretion rate onto the star becomes sufficiently low, and pre-main sequence phase starts. It takes another 2--4 Myr for photoevaporation by X-rays to create a gap in disk at $\sim 1$ au. Soon after the gap formation, gas of the inner disk accretes to the star, and disk becomes unobservable in NIR. If a pre-stellar core has higher angular velocity or lower central density, its evolution timescale is longer than typical one. Also, the lower X-ray luminosity of the central star is, the longer the evolution timescale is. In addition, we have estimated the disk fraction at a given stellar age, constructing the distribution functions of pre-stellar cores and X-ray luminosities of YSOs by using the observational data. As a result, the mean lifetime of protoplanetary disks is $t_{\rm life,m}\simeq 3.7$ Myr, which is consistent with the observational estimate. We have also found that the X-ray luminosity has the strongest impact on the disk lifetime in our model.

Finally, we discuss the effects of ignored processes, which may be the reason why our model slightly overestimates the disk fraction systematically. First, we ignore far ultraviolet (FUV) photons from the central star. 
Although the FUV photoevaporation rate in \citet{GDH09a} is comparable to the X-ray photoevaporation rate in \citet{Owe12a}, it is complex and still a work in progress \citep[see][for a more detailed review]{APA14a}.
The large scale magnetic field also seems to affect the disk lifetime, which is ignored in our model for simplicity. The angular momentum in a collapsing pre-stellar core is transported outward by magnetic braking during the infall phase \citep[e.g.,][]{MIM11a}. This makes disk radius smaller, which results in shorter disk lifetime. Also, the large scale magnetic field drives disk winds via turbulent thermal pressure \citep[e.g.,][]{SMI10a}, or magnetic centrifugal force \citep[e.g.][]{BP82a}. These winds also shorten the disk lifetime through the transport of angular momentum and mass. This effect of disk winds may be dominant process of disk dispersal if there are strong vertical magnetic fields, as discussed in \citet{Bai16a}. On the other hand, if we include the resistivity, some part of the disk becomes MRI inactive \citep[dead zone, e.g.][see Section \ref{sec:diskfraction}]{Gam96a,SM99a}. This probably makes disk lifetime longer, because the materials accumulate in the dead zone \citep{BHZ13a}. In addition, as discussed in Section \ref{sec:typical}, our treatment of stellar evolution is incomplete. In reality, $t_{\rm pms}$ should be shorter, which probably makes the disk lifetime longer. More realistic calculations including these effects remain as a future work.


\section*{Acknowledgements}

We thank Jaehan Bae, Chikako Yasui, and Shu-ichiro Inutsuka for useful comments. MK is supported by Grants-in-Aid from MEXT of Japan (Grant: 23244027). 









\appendix

\section{Estimation of cutoff radius}

Consider a disk with the following surface density profile,
\begin{equation}
 \Sigma(r) = \Sigma_0 \left({r\over r_{\rm cut}}\right)^{-p}\exp\left(-{r\over r_{\rm cut}}\right), \label{eq:sig_profile}
\end{equation}
where $r_{\rm cut}$ is the cutoff radius.
The enclosed disk mass inside $r$ is estimated to be 
\begin{eqnarray}
 M(r) &=& 2\pi \int_{r_{\rm in}}^r r' \Sigma(r') dr' \nonumber \\
&=& 2\pi r_{\rm cut}^2\Sigma_0 \gamma\left(-p+2,~r/r_{\rm cut}\right),\label{eq:mr}
\end{eqnarray}
where $\gamma(a,x)$ is the lower incomplete gamma function. The total disk mass is described as 
\begin{equation}
M_{\rm disk}=2\pi r_{\rm cut}^2 \Sigma_0 \Gamma(-p+2), \label{eq:md}
\end{equation}
where $\Gamma(a)$ is the gamma function. Dividing Equation (\ref{eq:mr}) of $r=r_{\rm cut}$ by (\ref{eq:md}), we obtain the cutoff radius from following equation,
\begin{equation}
 M(r_{\rm cut}) = M_{\rm disk} {\gamma(-p+2,1)\over \Gamma(-p+2)}. \label{eq:rcut}
\end{equation}
The disk profile is $\Sigma(r) \propto r^{-15/14}$ when the irradiation from the central star is the dominant heating source. In this case, $\gamma(-p+2,1)/\Gamma(-p+2)\simeq0.663$. Using this value and (\ref{eq:rcut}), we tabulate the resultant cutoff radius at $t=t_{\rm pms}$ in Table \ref{tab:results}. Independently, we get $r_{\rm cut}$ by directly fitting the simulation result by Equation (\ref{eq:sig_profile}), and it is found that these two estimates are consistent within a factor 1.2.



\bsp	
\label{lastpage}
\end{document}